\documentclass{ws-p8-50x6-00}

\begin{document}
%%% start MY title page %%%%%%%%%%%%%

%\begin{titlepage}
\thispagestyle{empty}

\begin{flushright}
IRB-TH-9/00\\
September, 2000
\end{flushright}

\vspace{2.0cm}

\begin{center}
\Large\bf Lifetime pattern of heavy hadrons
\vspace*{0.3truecm}
\end{center}

\vspace{1.8cm}

\begin{center}
\large B. Guberina, B. Meli\'{c} and
 H. \v Stefan\v ci\'c\\
{\sl \hspace*{0.5cm} 
Theoretical Physics Division, Rudjer Bo\v skovi\'c Institute,
\\
P.O.Box 180, HR-10002 Zagreb, Croatia\\[3pt]
E-mails: {\tt guberina@thphys.irb.hr,
         \tt melic@thphys.irb.hr,
         \tt shrvoje@thphys.irb.hr}}
\end{center}

\vspace{2.0cm}

\begin{center}
{\bf Abstract}\\[0.3cm]
\parbox{13cm}
{
We discuss the lifetime pattern of weakly decaying heavy hadrons. 
}
\end{center}

\vspace{2.5cm}

\begin{center}
{\sl Talk given by B. Meli\'c at
The 30th International Conference on High Energy Physics, 
July 27 - August 2, 2000, Osaka, Japan\\
To appear in the Proceedings}
\end{center}

\vbox{}
\thispagestyle{empty}
\newpage
\setcounter{page}{1}
\thispagestyle{empty}

%%% end MY title page %%%%%%%%%%%%%

\title{Lifetime pattern of heavy hadrons}

\author{B. Guberina, B. Meli\'c and H. \v Stefan\v ci\'c}

\address{Theoretical Physics Division, Rudjer Bo\v{s}kovi\'{c} Institute, \\
   P.O.Box 180, HR-10002 Zagreb, Croatia
\\E-mails: guberina@thphys.irb.hr, melic@thphys.irb.hr, shrvoje@thphys.irb.hr}

\maketitle

\abstracts{
We discuss the lifetime pattern of weakly decaying heavy hadrons. 
}
%In order to eliminate the model dependence of
%preasymptotic effects in inclusive heavy hadron decays,
%a connection between charmed and beauty sectors has to be
%made. Assuming isospin and heavy quark symmetry a set of relations
%connecting charm and beauty decay rates is obtained without
%invoking specific models.
%%A relation revealing universality in decays of different
%%heavy hadrons is established making possible to introduce some
%%order in otherwise rather intricate pattern of heavy hadron lifetimes.

\section{Introduction\label{sec:intro}}%1
%
%The inclusive decay of a weakly decaying heavy hadron $H_Q$ is 
%described   
%%inside the Heavy Quark Effective Theory (HQET) 
%by the well-know expression  
%as a sum of matrix elements of operators of increasing 
%dimensionality :
%\begin{eqnarray}
%\Gamma (H_{Q} \rightarrow f) &=& \frac{G_{\rm F}^2 m_Q^5}{192 \pi^3} |V|^2
%\frac{1}{2 M_{H_{b}}} \{ \nonumber \\
% & & \hspace*{-2.1cm} c_{\overline{Q}Q}^f
%\langle H_{Q}|\overline{Q}Q|H_{Q}\rangle
%\nonumber \\ & & 
%\hspace*{-2.1cm} + c_G^f \frac{ \langle H_{Q}|
%\overline{Q}g\sigma_{\mu\nu}G^{\mu\nu} Q
%|H_{Q}\rangle }{m_Q^2}  \nonumber \\
% & & \hspace*{-2.1cm} + \sum_i \left (c_{4q}^f \right )_i
%\frac{ \langle H_{Q}|\overline{Q} \Gamma_i Q \,\overline{q}
%\Gamma_i^{'} q |H_{Q}\rangle }{m_Q^3}
%+ ...\}, 
%\label{eq:main}
%\end{eqnarray}
%where the first matrix element 
%$\langle H_{Q}|\overline{Q}Q|H_{Q}\rangle$ has to be 
%systematically further expanded in $1/m_Q$. Corrections of the order 
%$1/m_Q$ do not appear in the above expression and 
%leading corrections are suppressed by $1/m_Q^2$.  
%Further corrections of the order $1/m_Q^3$ are given by the matrix elements of 
%four-quark operators and are known as the positive/negative 
%Pauli interference and weak exchange effects. 
A lot of physical observables in heavy-quark decays are described using 
the inverse heavy-quark mass expansion in terms of a few basic quantities, 
i.e., quark masses and hadronic expectation values of several local 
operators \cite{bigi}. 
The remarkable fact that the expansion is 
applicable even to the
the charmed case enables one to connect charmed and 
beauty sectors. 
%The remarkable fact that the expansion is applicable even to 
%charmed-hadron decays,  
%(inspite of the largeness of the expansion 
%parameter $\sim 0.5$), 
%indicates the same driving mechanism in both 
%charmed and beauty decays, which in turn enables one to search for 
%possible connections between these decays. 
In this presentation
\footnote{
Talk given by B. Meli\'c at ICHEP2000, July 27 -
August 2, Osaka, Japan}
we relate lifetimes of heavy hadrons and obtain some interesting 
predictions. 

\begin{figure}%1
\epsfxsize250pt
\figurebox{120pt}{160pt}{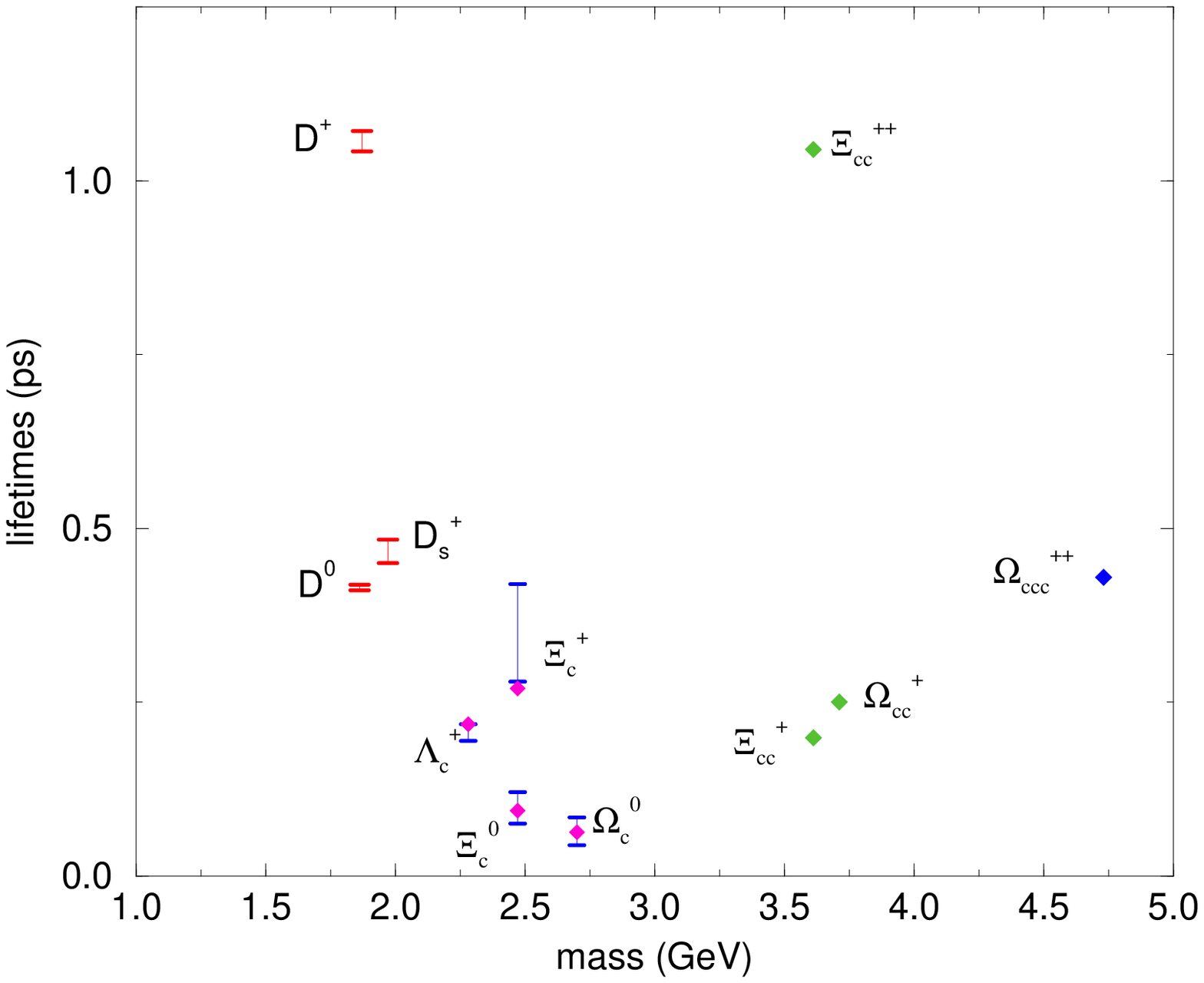}
\caption{Lifetimes of weakly decaying charmed hadrons. Diamonds denote 
theoretical predictions.}
\label{fig:charm}
\end{figure}

Large lifetime differences between charmed hadrons, shown in 
Fig.~\ref{fig:charm}, 
cannot be explained by taking just ${\cal O}(1/m_c^2)$ corrections into 
account. The diversity among charmed hadron lifetimes is attributed 
to the effects of four-quark operator (${\cal O}^{4q}$) contributions 
of the order of $1/m_c^3$. 
For charmed mesons \cite{cmesons} and baryons \cite{cbaryons}, 
the theoretical findings have been also confirmed 
by experiment. Analogously, a significant spread of doubly charmed 
baryon lifetimes is predicted \cite{ccbaryons}, Fig.~\ref{fig:charm}.  

For beauty hadron decays there is a rapid convergence of the $1/m_b$ 
expansion and the leading ${\cal O}(1/m_b^2)$ corrections 
introduce a difference of just $2-3\%$ between lifetimes of beauty hadrons. 
But, owing to the peculiarity of the $1/m_Q$ expansion, exhibited also in charmed 
decays, there is still 
some place for ${\cal O}(1/m_Q^3)$ operators to play a significant role in 
beauty decays. In 
a recent work \cite{bbaryons} we have found an enhancement of 
the ${\cal O}^{4q}$ 
contributions in beauty baryon decays. Such an enhancement brings the 
$\tau(\Lambda_b)/\tau(B)$ ratio much closer to the experimental value 
compared with 
the standard nonrelativistic-model estimation and 
predicts a much larger spread among beauty-baryon lifetimes, 
Fig.~\ref{fig:beauty}. 

\section{Connecting charm and beauty}%1

As in the calculation of the ${\cal O}^{4q}$ contributions one relies on the 
questionable nonrelativistic models, it is a challenge to search for some 
model-independent determination of such contributions. Having in mind 
the same 
formalism applied in the treatment of both charmed and beauty decays, 
and large effects which the ${\cal O}^{4q}$ 
contributions exhibit in the lifetimes of charmed as well as beauty hadrons,  
discussed in Sec.~\ref{sec:intro}, 
we try to perform a model-independent analysis searching for an explicit 
connection between charmed and beauty sectors.  

\begin{figure}%1
\epsfxsize250pt
\figurebox{120pt}{160pt}{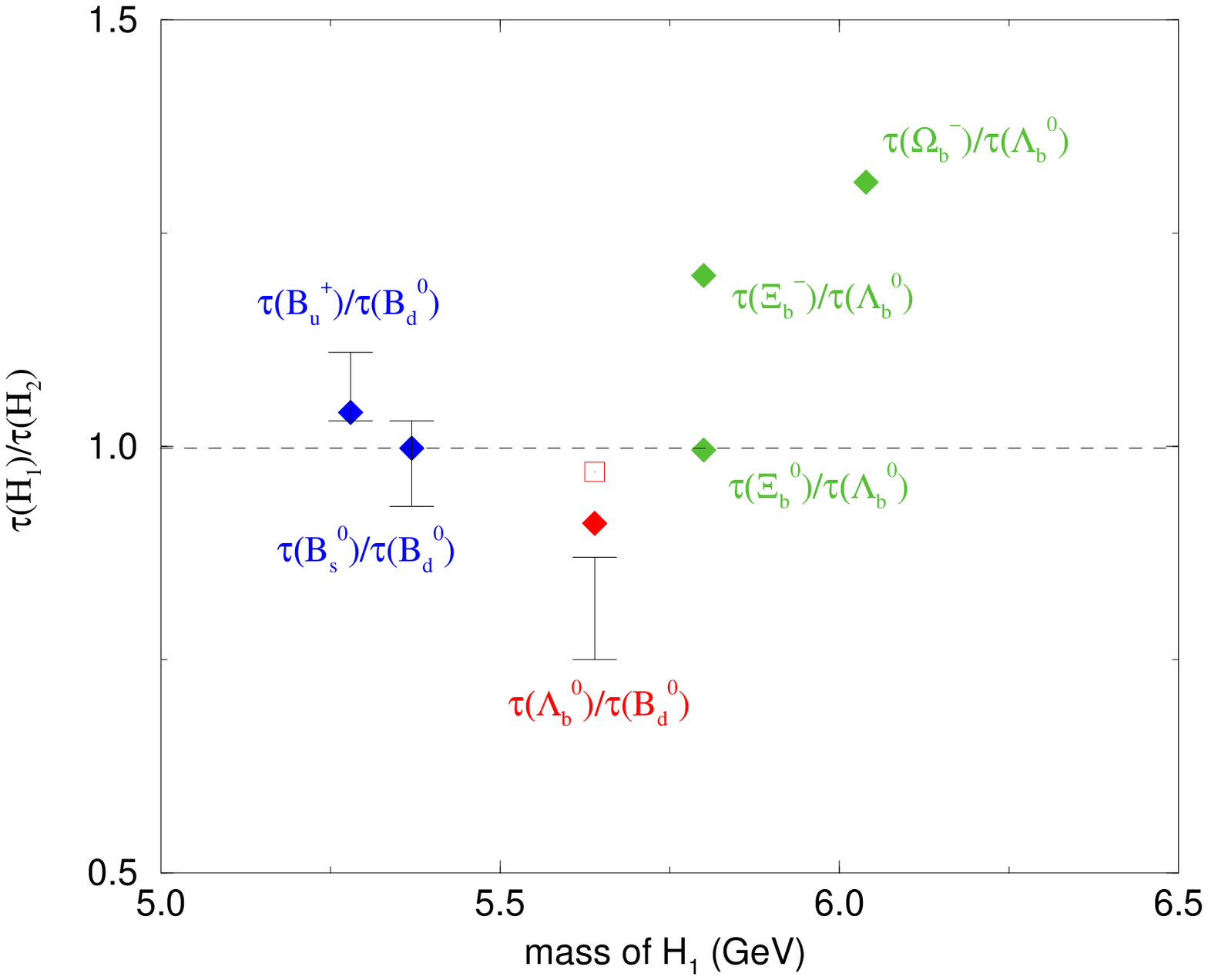}
\caption{Lifetimes of weakly decaying beauty hadrons. Diamonds denote 
theoretical predictions. The square stands for the standard 
nonrelativistic model prediction of the $\tau(\Lambda_b)/\tau(B)$ ratio.}
\label{fig:beauty}
\end{figure}

Similar ideas have already been applied, first by Voloshin \cite{voloshin},  
to obtain the lifetime differences between beauty hyperons, and then by us 
\cite{bbaryons}, where we have obtained, in a moderate model-dependent 
analysis, an enhancement of four-quark contributions for beauty baryons with 
predictions shown in Fig.~\ref{fig:beauty}. 
We group heavy hadrons 
exhibiting the same type of ${\cal O}^{4q}$ contributions in pairs 
and consider their decay-rate differences applying the SU(2) isospin symmetry 
and the heavy-quark symmetry (HQS) \cite{diff}. 
%Here we apply the SU(2) isospin symmetry and the heavy quark symmetry (HQS) 
%and connect heavy hadrons affected with the same type of the four-quark 
%contribution. 
Cabibbo suppressed modes and 
mass corrections in final states are neglected. 
We start our investigation by treating different 
heavy-hadron sectors separately. 
\subsection{Heavy mesons}\label{subsec:mesons}%1.1
The first pair of considered mesons form $D^+$ and $B^-$. In decays of both 
of these mesons the negative Pauli interference occurs. The second pair 
of particles having the effects of a weak exchange form $D^0$ and $B^0$. The 
idea is to combine decay rates in such a manner that the effects of 
the ${\cal O}^{4q}$ contributions are isolated:
\begin{eqnarray*}
\Gamma(D^+) - \Gamma(D^0) = \frac{G_F^2 m_c^2}{4 \pi} |V_{cs}|^2 |V_{ud}|^2
\times \\
\left [ \langle D^+|{\cal L}_{\rm PI}^{cd}| D^+ \rangle -
\langle D^0|{\cal L}_{\rm exc}^{cu}| D^0 \rangle \right ] \nonumber \\
\Gamma(B^-) - \Gamma(B^0) = \frac{G_F^2 m_b^2}{4 \pi} |V_{cb}|^2 |V_{ud}|^2
\times \\
\left [ \langle B^-|{\cal L}_{\rm PI}^{bu}| B^- \rangle -
\langle B^0|{\cal L}_{\rm exc}^{bd}| B^0 \rangle \right ] \,. 
\end{eqnarray*}
${\cal L}$'s are parts of the weak Lagrangian involving Cabibbo-leading 
nonleptonic as well as semileptonic parts. 

Assuming the isospin symmetry in the heavy-quark limit, 
the ${\cal O}^{4q}$ 
contributions get reduced and we obtain the 
following relation:
\begin{eqnarray}
R^{BD} &=& \frac{\Gamma(B^-) - \Gamma(B^0)}{\Gamma(D^+)- \Gamma(D^0)} =  
\nonumber \nonumber\\  &=& 
\frac{m_b^2}{m_c^2} \frac{|V_{cb}|^2}{|V_{cs}|^2}\left [1 +
{\cal O}(1/m_{b,c}) \right ] \, . 
\label{eq:mesR}
\end{eqnarray}
Starting from this expression, we may check the standard formalism of 
inclusive decays, expressed by Eq.~(\ref{eq:mesR}), 
against experimental data on 
heavy-meson lifetimes \cite{PDG}. The results are 
\begin{eqnarray*}
R^{BD} &=& 0.020 \pm 0.007 \, , \\
R^{BD}_{\rm exp} &=& 0.030 \pm 0.011 \, , 
\end{eqnarray*}
and they are consistent within errors. 
\subsection{Heavy baryons}\label{subsec:baryons}%1.1
In baryon decays we find $\Xi_c^+$ and $\Xi_b^-$ experiencing the 
negative interference, and $\Xi_c^0$ and $\Xi_b^0$ exhibiting the 
weak exchange. There are also different nonleptonic and semileptonic 
contributions from the operators involving s-quark, but they cancel 
in the decay-rate differences: 
\begin{eqnarray*}
\Gamma(\Xi_c^+) - \Gamma(\Xi^0_c) = \frac{G_F^2 m_c^2}{4 \pi} |V_{cs}|^2
|V_{ud}|^2
\times \\
\left [ \langle \Xi^+_c|{\cal L}_{\rm PI}^{cu}| \Xi^+_c \rangle -
\langle \Xi^0_c|{\cal L}_{\rm exc}^{cd}| \Xi^0_c \rangle \right ] \\
\Gamma(\Xi_b^-) - \Gamma(\Xi^0_b) = \frac{G_F^2 m_b^2}{4 \pi}
|V_{cb}|^2 |V_{ud}|^2
\times \\
\left [ \langle \Xi^-_b|{\cal L}_{\rm PI}^{bd}| \Xi^-_b \rangle -
\langle \Xi^0_b|{\cal L}_{\rm exc}^{bu}| \Xi^0_b \rangle \right ] \, . 
\end{eqnarray*}
Applying SU(2) symmetry and HQS as before, we obtain  to 
order $1/m_{c,b}$
\begin{eqnarray}
R^{bc} &=& \frac{\Gamma(\Xi^-_b) - \Gamma(\Xi^0_b)}
{\Gamma(\Xi^+_c)- \Gamma(\Xi^0_c)} 
= \frac{m_b^2}{m_c^2} \frac{|V_{cb}|^2}{|V_{cs}|^2}
%\left [ 1  + {\cal O}(1/m_{b,c}) \right ] \, . 
\label{eq:barR}
\end{eqnarray}
This relation can serve as a test of the model-dependent predictions, 
presented in Figs.1 and 2. If we calculate $R^{bc}$ using 
model-dependent approach \cite{cbaryons,bbaryons} 
consistently with approximations 
made in this analysis, we obtain a difference of $12\%$ compared 
with the prediction obtained from  
Eq.~(\ref{eq:barR}). 
By performing the complete calculation with the mass 
corrections and Cabibbo-suppressed modes included, we can judge 
the order of neglected corrections to be less than $10\%$. 

The relation (\ref{eq:barR}) enables us to obtain a prediction for the lifetime 
difference between beauty hyperons, using measured lifetimes of 
singly-charmed baryons $\Xi_c^+$ and $\Xi_c^0$ \cite{cbaryons}:
\begin{eqnarray*}
\Gamma(\Xi_b^-) - \Gamma(\Xi_b^0) = -(0.14 \pm 0.06) \; {\rm ps^{-1}}. 
\end{eqnarray*}
The prediction can be compared with the values from \cite{voloshin} and 
\cite{bbaryons}, where $SU(3)_f$ symmetry and HQS were used. All results 
appear to be consistent with each other. 
%\\
%-to be compared with\\
%i) model-independent prediction (${\rm SU(3)_f}$ and HQS)\\
%\hspace*{1cm} $\Gamma(\Xi_b^-) - \Gamma(\Xi_b^0)
%= -(0.11 \pm 0.03) \; {\rm ps^{-1}}$ \\
%ii) moderate model-dependent prediction \\(figure !)\\
%\hspace*{1cm} $\Gamma(\Xi_b^-) - \Gamma(\Xi_b^0)
%= -(0.094) \; {\rm ps^{-1}}$
%\end{minipage}
%}
%
\subsection{Doubly-heavy baryons}\label{subsec:doublybaryons}%1.1
Finally, a similar procedure applies to doubly-heavy baryons. 
The obtained expression now relates doubly beauty baryons with 
doubly-charmed baryons to order $1/m_{c,b}$:
\begin{eqnarray}
R^{bbcc} &=& \frac{\Gamma(\Xi^-_{bb}) - \Gamma(\Xi^0_{bb})}
{\Gamma(\Xi^{++}_{cc})- \Gamma(\Xi^+_{cc})} 
= \frac{m_b^2}{m_c^2} \frac{|V_{cb}|^2}{|V_{cs}|^2}
%\left [1 + {\cal O}(1/m_{b,c}) \right ]. 
\label{eq:doublyR}
\end{eqnarray}
Unfortunately, there is still no experimental evidence for 
doubly-heavy baryon lifetimes to check the above relation. 
However, we can use it  
to calculate the splitting among doubly-beauty hyperons $\Xi_{bb}^-$ 
and $\Xi_{bb}^0$, by taking existing theoretical predictions for 
doubly-charmed lifetimes \cite{ccbaryons}. The prediction for 
the doubly-beauty lifetime spread is
\begin{eqnarray*}
\Gamma(\Xi_{bb}^-) - \Gamma(\Xi_{bb}^0) = - 0.073 \, {\rm ps^{-1}}. 
\end{eqnarray*}
\section{Conclusions}%1

By inspection of all three relations (\ref{eq:mesR}), (\ref{eq:barR}) and 
(\ref{eq:doublyR}) we can see 
that there exists a universal behavior in decays of heavy hadrons 
summarized by the expression: 
%\begin{table*}[h]
\begin{eqnarray*}%1
\frac{\Gamma(B^-) - \Gamma(B^0)}{\Gamma(D^+)- \Gamma(D^0)} &=& 
\frac{\Gamma(\Xi^-_b) - \Gamma(\Xi^0_b)}
{\Gamma(\Xi^+_c)- \Gamma(\Xi^0_c)} = \nonumber \\
= \frac{\Gamma(\Xi^-_{bb}) - \Gamma(\Xi^0_{bb})}
{\Gamma(\Xi^{++}_{cc})- \Gamma(\Xi^+_{cc})}
&=&\frac{m_b^2}{m_c^2} \frac{|V_{cb}|^2}{|V_{cs}|^2} \,. 
\label{eq:summ}
\end{eqnarray*}
%\end{table*}
%
This relation connects all sectors of weakly decaying heavy hadrons that are 
usually treated separately: mesons and baryons, 
charmed and beauty particles, and 
brings some order in the otherwise rather intricate pattern of heavy-hadron 
lifetimes. The predictions we have obtained, although burdened with some 
approximations, if experimentally confirmed 
would indicate that four-quark operators can account for the greatest part 
of the decay rate differences among heavy hadrons. 

%\section*{Acknowledgments}
\vspace*{0.5cm}
\noindent
This work was supported by the Ministry of
Science and Technology of the Republic of Croatia under the contract 
No. 00980102.

\end{document}